\documentclass[slac_one]{revtex4}
\usepackage{graphicx}
\usepackage{fancyhdr}
\newcommand{\ampl}       {{\cal{A}}
}

\newcommand{\ds}        {\ensuremath{{D_{s}^{-}}}}

\newcommand{\dmd}       {\ensuremath{{\Delta {m}_{d}}}}
\newcommand{\dms}       {\ensuremath{{\Delta {m}_{s}}}}

\newcommand{\bs}        {\ensuremath{{B_{s}^{0}}}}

\newcommand{\dstophipi}   {\ensuremath{{D_{s}^{-} \rightarrow \phi \pi^{-}}}}
\newcommand{\dstokstk}   {\ensuremath{{D_{s}^{-} \rightarrow K^{\ast 0} K^-}}}
\newcommand{\dstokshk}   {\ensuremath{{D_{s}^{-} \rightarrow K_{S} K^-}}}
\newcommand{\bstodsmunux} {\ensuremath{{B_{s}^{0} \to D_{s}^{-} \mu^+ \nu X}}}
\newcommand{\bstodse} {\ensuremath{{B_{s}^{0} \to D_{s}^{-} e^+ \nu X}}}
\newcommand{\bstodspi} {\ensuremath{{B_{s}^{0} \to D_{s}^{-} \pi^{+} + X}}}

\newcommand{\GVc}       {\ensuremath{{{\rm GeV}/c}}}
\newcommand{\GVcc}       {\ensuremath{{{\rm GeV}/c^2}}}

\newcommand{\kstk}       {\ensuremath{K^{\ast 0} \rightarrow K^+ \pi^-}}
\newcommand{\kst}       {\ensuremath{K^{\ast 0}}}
\newcommand{\merit}     {\ensuremath{\varepsilon {\cal{D}}^2}}

\begin{document}

\title{$B_s^0$ mixing at D0 experiment} 

%

\author{T. Moulik (for the D0 collaboration)}
\affiliation{Department of Physics, University of Kansas, Lawrence, KS 66045, USA}

\begin{abstract}
In this report, we present a preliminary measurement of the 
$\bs$ mixing parameter using samples of four partially reconstructed
semileptonic $B_s^0$ decays and one fully reconstructed hadronic
decay mode, corresponding to approximately 2.4 $fb^{-1}$ of integrated 
luminosity. We perform an unbinned likelihood fit to
the proper decay length and obtain $\Delta m_s = 18.53 \pm 0.93 (\rm stat.)
\pm 0.30 ({\rm syst.}) ~ps^{-1}$.
\end{abstract}

\maketitle

\thispagestyle{fancy}

\section{INTRODUCTION}\label{intro}
Particle-antiparticle oscillations are observed and well established
in the $B_d^0$ system. The mass difference $\dmd$ is measured to be
$\dmd = 0.507 \pm 0.005$ ps$^{-1}$ \cite{pdg}. $B_s^0$ mesons are
known to oscillate with a high frequency according to standard model
predictions. Observing the oscillations in the $B_s^0$ system 
has been an important focus of the B physics 
program at both the D0 and the CDF experiments at Tevatron.
D0 reported direct limits on the $\bs$ mixing parameter
$\dms$ \cite{bsmixd0} using the $\bstodsmunux$, $\dstophipi$ decay mode
\footnote{Charge conjugated states are implied throughout the text}. CDF reported a measurement of this parameter
exceeding 5 $\sigma$ significance \cite{bsmixcdf}. 
The measurement of this parameter
is an important test of the CKM (Cabibbo Kobayashi Maskawa) formalism of 
the standard model, and combining it with a measurement of $\dmd$ allows us to
reduce the error on $V_{td}$ and constrain one side of the CKM triangle.
This report describes the measurement of the $B_s^0$ mixing parameter
at the D0 experiement using 2.4 $fb^{-1}$ of data with 3 $\sigma$
significance.

We use the central tracker, muon chambers and calorimeters to
reconstruct the $B$ decays. Details of the detector can be found elsewhere \cite{d0det}. 
We use a single inclusive muon trigger or a di-muon trigger to 
accumulate samples for $\bs$ mixing studies. In the case of 
di-muon trigger the other muon acts as the tag muon used to
identify the flavor of the $B$ meson which we discuss later.
The trigger requires a good muon identified by the muon chamber
with a matching track in the central tracker in the pseudo-rapidity
range of $|\eta| < 2.0$. The triggers use
$p_T$ cuts between $3 - 5$ $\GVc$ and the trigger is prescaled or turned 
off depending on the luminosity. Hence for the $\bstodse$ decay mode,
and the hadronic decay mode, we are using a tagged sample with the 
muon acting as a tag. 

\section{$B_s^0$ DECAYS SAMPLE SELECTION AND RECONSTRUCTION}
D0 reported direct limits on $B_s^0$ oscillations using the decay mode
$\bstodsmunux$ decays with $\dstophipi$ \cite{bsmixd0}. In this report, we present 
a preliminary measurement of the $B_s^0$ oscillations \cite{bsmixd0meas}, using four additional decay
modes. We use three additional semileptonic decay modes namely $\bstodse$, with $\dstophipi$ and $\bstodsmunux$ decays with 
$\dstokstk$, $K^{\ast 0} \rightarrow K^+ \pi^-$ (The  
$K^{\ast 0}$ and the $\phi$ candidates are required to be consistent
with known mass and width \cite{pdg} of these two resonances) and
$\dstokshk$.
We also use a fully reconstructed hadronic mode, namely, $\bstodspi$ decay,
with $\dstophipi$.

$\bs$ decays are identified in their semileptonic modes, both 
in muon and electron mode. Muons are required to have $p_T > 1.5 \GVc$,
and to be in the pseudo-rapidity region of $|\eta| < 2.0$. Electrons are 
required to have $p_T > 2.0$ $\GVc$ and are identified in a pseudo-rapidity
region of $|\eta| < 1$.  The $D_s^-$ and $B_s^0$ decay products are constrained to originate
from a common vertex and the $B_s^0$ and $D_s^{-}$ decay vertices
are required to be significantly displaced from the $p\bar{p}$ collision
vertex. 

\begin{figure}[h]
\includegraphics[width=4in]{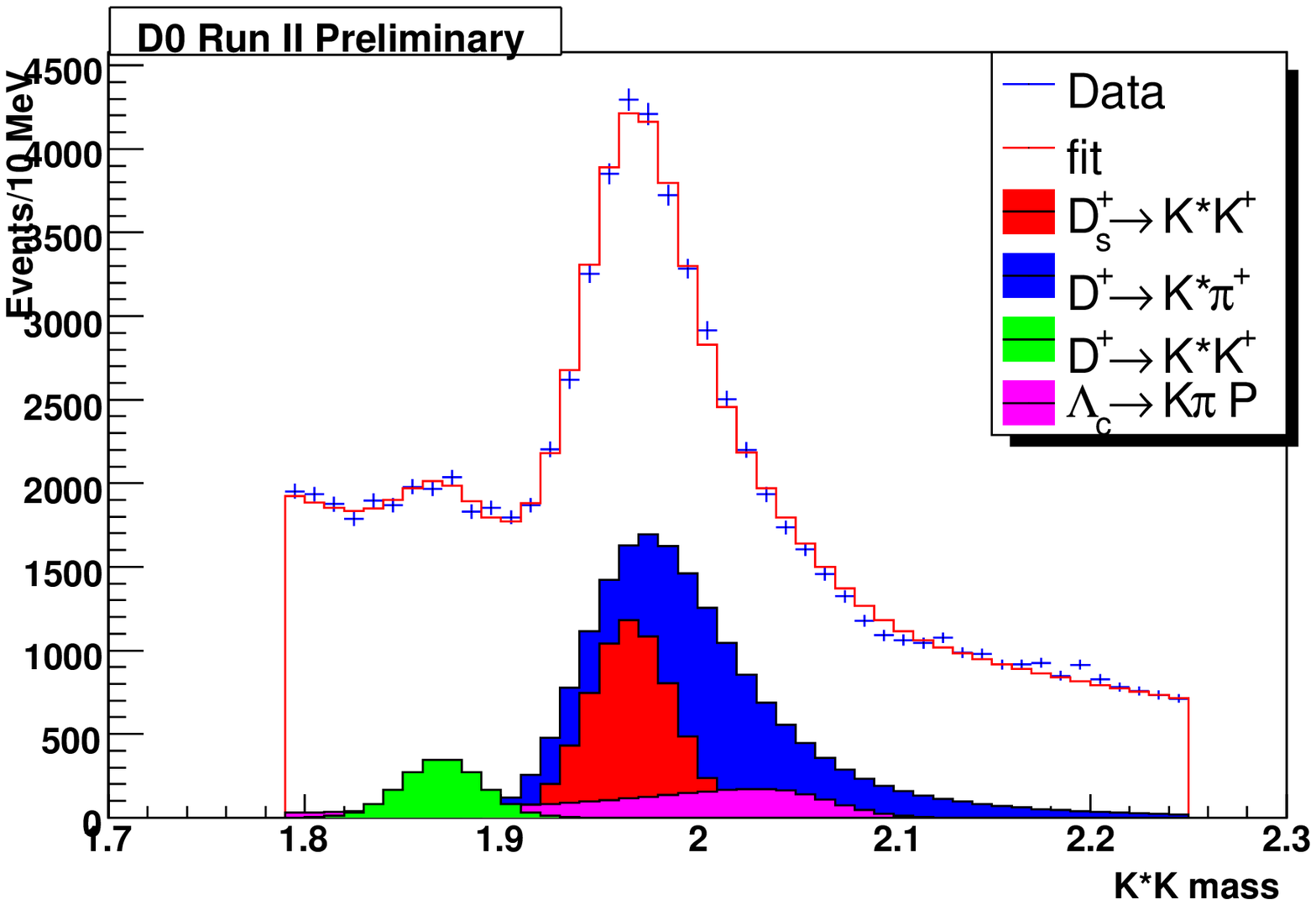}
\caption{The $M_{(K\pi)K}$ invariant mass distribution where $D_s \rightarrow \kst K$ for $\bstodsmunux$ decay mode}
\label{fig:mukstk}
\end{figure}

The $\kstk$ decay mode requires special treatment on account of large reflections,
as both real physics processes and combinatorial background contribute
to the signal peak. The peak is comprised of the signal mode, $\dstokstk$, $\kstk$,
the physics processes, $D^+ \rightarrow K^- \pi^+ \pi^+$ or $D^+ \rightarrow K^{\ast 0} \pi^+ (K^{\ast 0} \rightarrow K^+ \pi^-)$,
$\Lambda_c^{+} \rightarrow K^{+} \pi^{-} p^+$, $D^{+} \rightarrow K^{\ast 0} K^+ (\kstk)$ (Cabibbo suppressed) and
combinatorial background. We fit for these contributions in an unbinned likelihood
fit. Fig. \ref{fig:mukstk} shows the fit to the
mass distribution of the $M_{K \pi K}$ system with the individual contributions
superimposed. 

\section{FLAVOR TAGGING} \label{sec:tag}
The flavor of the initial state of the $\bs$ is determined using a likelihood ratio method, 
based on the properties of the other b-hadron in the event (opposite side tagging) and
the properties of the particles produced in assosciation with the reconstructed
$B_s^0$ meson (same side tagging) and we then combine the OST and the SST taggers.
The performance of any  tagger is characterized by its efficiency defined as 
$\varepsilon = N_{tag}/N_{tot}$ (where $N_{tag}$ is the number of tagged $B_s^0$ mesons 
and $N_{tot}$ is the total reconstructed $B_s^0$ mesons), and the dilution ${\cal{D}}$ which is
defined as ${\cal{D}} = \frac{N_{RS} - N_{WS}}{N_{tag}}$ (where $N_{RS}$ and $N_{WS}$ are 
the right-sign and wrong-sign tags respectively). 

The OST is calibrated using $B \rightarrow \mu+ \nu D^{\ast -}$ data events,
and we obtain the dilution ${\cal{D}}$ as a function of a tag variable 
$d$ (whose sign indicates a $b$ or $\bar{b}$, and its value indicates the ``$b$''-ness of the tag), 
to provide an event-by-event ``predicted'' dilution which 
is used in the unbinned likelihood fit described in section \ref{sec:likelihood}. 
More details on the  development of the OST and results can be found in \cite{bdmixd0}.
The total OST effective tagging power of $\merit = (2.48 \pm 0.21 ^{+0.08}_{-0.06})\%$.

\begin{figure}[h]
\includegraphics[width=3.0in]{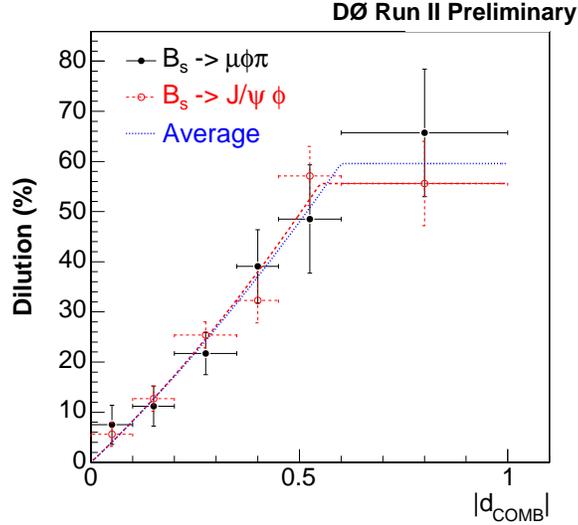}
\caption{The Combined Tagger dilution calibration}
\label{fig:combtag}
\end{figure}

The combined tagger is calibrated using decay modes, $B_s^0 \rightarrow J/\psi \phi$ (using
both data and MC) and $B_s^0 \rightarrow \mu D_{s} (\phi \pi)$ (using MC only) 
(See Fig. \ref{fig:combtag}). The total combined tagging power
is $\merit = 4.49 \pm 0.88 \%$. More details on the SST and the combined
tagger can be found in \cite{combtag}.

The SST is used by $\bstodsmunux$, with $\dstophipi$ decay mode only. All the
other decay modes use the OST. 

\section{UNBINNED LIKELIHOOD FIT AND AMPLITUDE SCAN}\label{sec:likelihood}

\subsection{Visible Proper decay length and scale factor}
In semileptonic decays the proper time gets smeared due to the presence of
neutrino. To take this into account we introduce
a K factor estimated from Monte Carlo (MC) simulation. It is defined as
$K = p_{T}(l D_s)/p_{T}(B)$. The real proper decay length 
is related to the measured or visible proper decay length ($l_{M}$), by the
relation $ct(B_s^0) = l_{M} K$ where $l_{M} = M(B_s^0) \cdot (L_T)/(p_T(l D_s))$ is the
measured visible proper decay length ($l_{M}$). $L_T$ is the distance from the
primary vertex to the $B_s^0$ decay vertex in the transverse plane projected
onto the $l D_s$ momentum, and $M(B_s^0)$ is the mass of the $B_s^0$ meson as obtained 
from PDG \cite{pdg}.

The $l_{M}$ uncertainty is determined by the vertex fitting procedure, track parameters,
and track parameter uncertainties. To account for any imperfections in modeling
of detector uncertainties, we use scale factors that are dependent on the 
topology of hits in the silicon micro-strip tracker to correct the $l_M$ uncertainty. 
These scale factor corrections were obtained from QCD data and MC samples.

\subsection{Likelihood Fit}
An unbinned likelihood fit is used to describe and fit for the
$\bs$ oscillation. All flavor tagged events with 
$1.72 < M (D_{s}) < 2.22$ $\GVcc$ are used in the fit.
The probability for oscillated (osc) and non-oscillated (nos) 
$\bs$ decays, as a function of true proper decay length,
$x$, can be writtten as 

\begin{equation}
p_{s}^{nos}(x) = \frac{K}{c \tau_{B_s}} \cdot e^{-\frac{K}{c \tau_{B_s}}} \cdot 0.5 \cdot
(1 + {\cal{D}} \cos(\Delta m_{s} \cdot Kx/c))
\end{equation}
\begin{equation}
p_{s}^{osc}(x) = \frac{K}{c \tau_{B_s}} \cdot e^{-\frac{K}{c \tau_{B_s}}} \cdot 0.5 \cdot
(1 - {\cal{D}} \cos(\Delta m_{s} \cdot Kx/c))
\end{equation}

The sample is mostly composed of $\bs$ decays with some contributions
coming from $B_d$ and $B_u$ mesons also. Similar equations would hold for these
decay modes with the oscillatory term getting modified appropriately.

Taking into account momentum uncertainty, and convoluting over detector
resolution, the probability for the $j$-th decay mode,
as a function of the measured proper decay length, can be written as below :
 
\begin{equation}
p_{j}^{nos/osc} (x^{M}) = \sum_{j} f_{j}(K) dK  \epsilon_{j}(x^M) \int dx ~g(x-x^M,x)  p_{j}^{nos/osc}(x,K)
\end{equation}

$f_j(K)$ is the K-factor distribution and we sum over the $K-$ factor bins.
$\epsilon_{j} (x^{M})$ is the reconstruction 
efficiency as a function of lifetime cuts. 
The fraction of each decay mode contribution is calculated from Monte Carlo. 
A similar probablity equation holds for the combinatorial background ($p_{bg}$). 
The background contribution comes from prompt decays with the 
$\ds \mu(e)$ vertex coinciding with the primary vertex, background with 
fake vertices distributed around the primary vertex, and long lived background.

The likelihood can then be written as follows :
\begin{equation}
{\cal{L}} = \prod_{i=1}^{N} ((1-f_{sig}) p_{i,bg} + f_{sig} p_{i,sig})
\end{equation}.

The amplitude method was first proposed elsewhere \cite{amplscan}.
It involves modifying the likelihood term by introducing an amplitude term
${\cal{A}}$ in front of the oscillatory cosine term, such that,

\begin{equation}
{\cal{L}} \propto 1 \pm {\cal {A ~D}} \cos (\Delta m_s t)
\end{equation}

We minimize $-2 ln {\cal L}$ fitting for the parameter ${\cal{A}}$ while ${\cal{D}}$ 
is known and $\Delta m_s$ is varied. The value of $\dms$ where $\ampl$ is consistent
with 1 and inconsistent with 0 then gives the measurement of the $\dms$ parameter.
All values of $\dms$ for which $\ampl + 1.645 ~\sigma_{\ampl} < 1$ 
are excluded at $95 \%$ confidence level. The sensitivity of the
mixing measurement is defined as the $\dms$ value for which 
$1.645 ~\sigma_{\ampl} = 1$. 

\begin{figure}[h]
\parbox{0.5\textwidth}{
\includegraphics[width=3in]{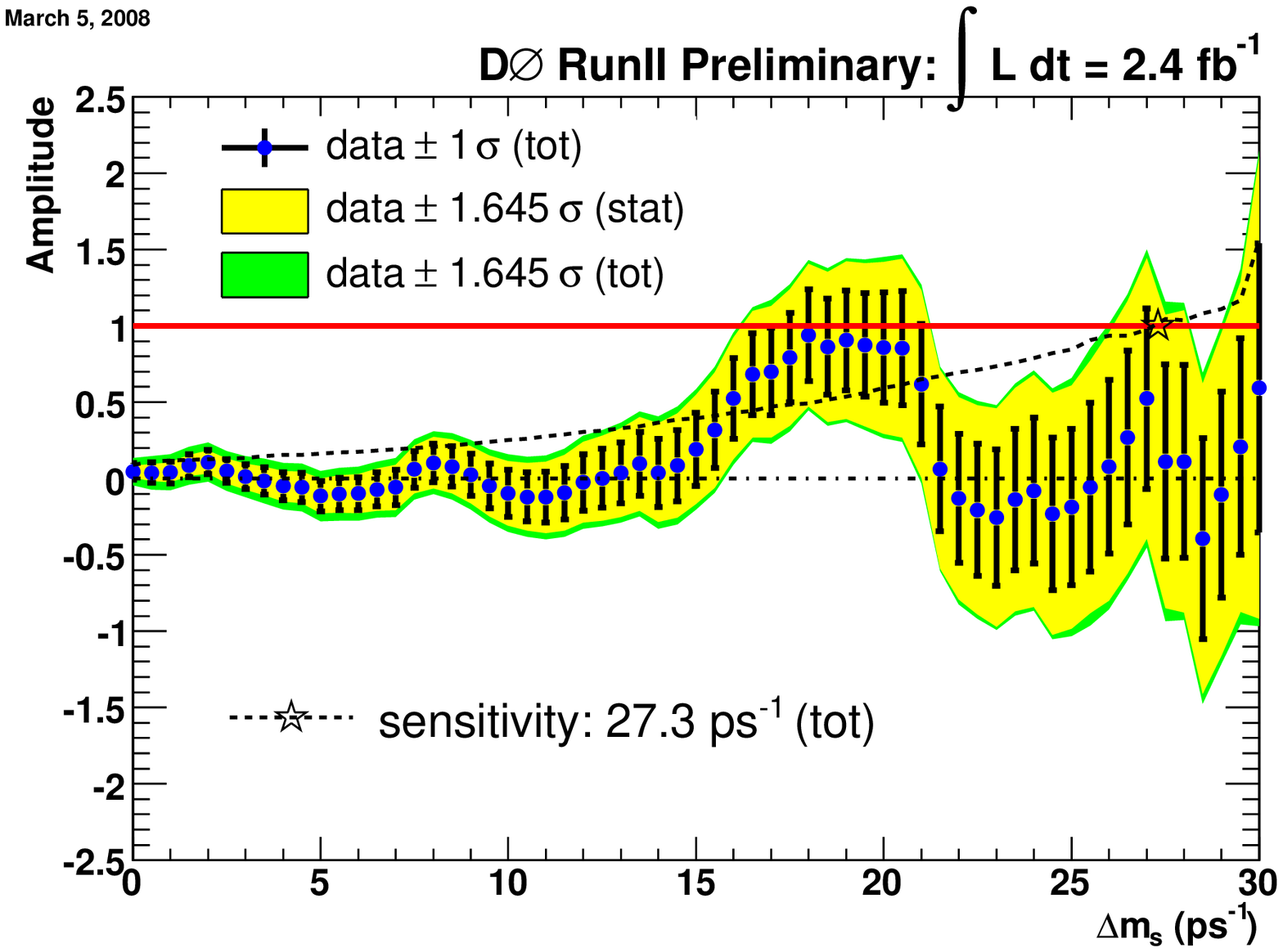}
}
\parbox{0.45\textwidth}{
\includegraphics[width=3in]{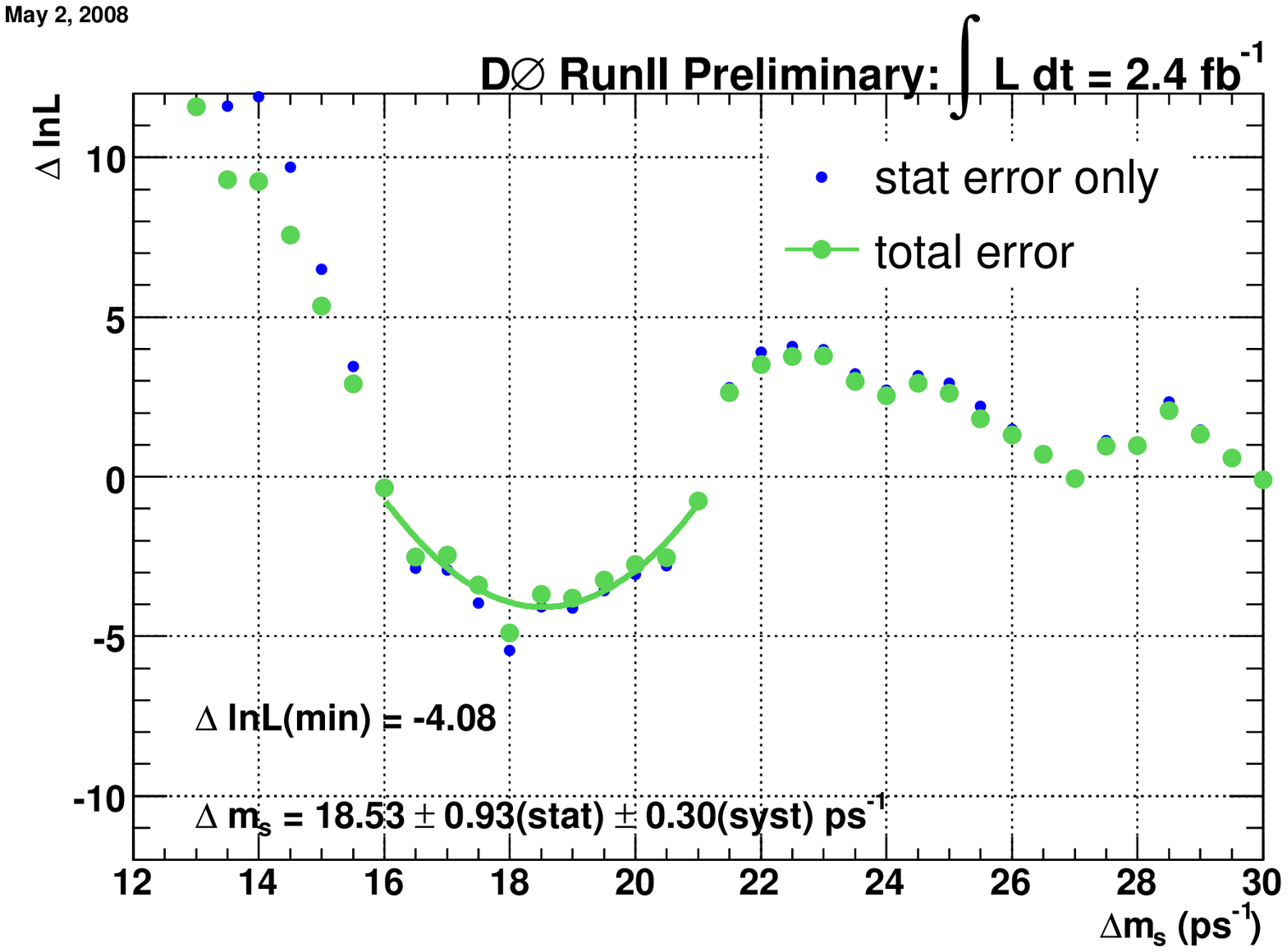}
}
\caption{The combined amplitude and likelihood scan.}
\label{fig:combampscan}
\end{figure}

The combined amplitude scan and the difference between log-likelihood 
at minimum and elsewhere ($\Delta {\rm ln} {\cal L}$), as a function 
of $\Delta m_s$, can be seen in Fig. \ref{fig:combampscan}. $\Delta {\rm ln} {\cal L}$
is derived from the amplitude scan. More details can be found in \cite{amplscan}.

From a parabolic fit around the minimum of
the $\Delta {\rm ln} {\cal L}$ distribution, 
we obtain $\Delta m_s = 18.53$ with statistical error of $0.93 ~ps^{-1}$. The 
systematic error is estimated as $0.3 ~ps^{-1}$, thus giving us a preliminary 
measurement of $\Delta m_s = 18.53 \pm 0.98 ~ps^{-1} \rm ~(stat. + syst.)$. 
From the depth of the likelihood minima, we calculate an overall significance of this 
measurement as $2.9 ~\sigma$.

\end{document}